\documentclass[11pt,
onecolumn,
 aip,
 jmp,%
 showkeys,
 amsmath,amssymb,
preprint,%
]{revtex4-1}

\usepackage[breaklinks=true,colorlinks=true,linkcolor=blue,urlcolor=blue,citecolor=blue]{hyperref}
\usepackage{graphicx}%
\usepackage{graphics}%
\usepackage{epstopdf}
\usepackage{dcolumn}%
\usepackage{soul,color}
\usepackage{bm}
\usepackage[mathlines]{lineno}
\expandafter\ifx\csname package@font\endcsname\relax\else
 \expandafter\expandafter
 \expandafter\usepackage
 \expandafter\expandafter
 \expandafter{\csname package@font\endcsname}%
\fi
\begin{document}
\abovedisplayskip=3pt
\belowdisplayskip=3pt
\abovedisplayshortskip=2pt
\belowdisplayshortskip=2pt


\title{\Large{AC-driven microwave loss modulation in a fluxonic metamaterial}}
\author{Oleksandr V.~Dobrovolskiy}
    \affiliation{Physikalisches Institut, Goethe University, 60438 Frankfurt am Main, Germany}
    \affiliation{Physics Department, V. Karazin National University, 61077 Kharkiv, Ukraine}
\author{Michael Huth}
    \affiliation{Physikalisches Institut, Goethe University, 60438 Frankfurt am Main, Germany}
\author{Valerij A. Shklovskij}
    \affiliation{Physics Department, V. Karazin National University, 61077 Kharkiv, Ukraine}
\date{\today}

\begin{abstract}
We introduce a fluxonic metamaterial on the basis of nanopatterned superconducting Nb microstrips and employ it for modulation and synthesis of quantized loss levels in the lower GHz range by a sine-wave quasistatic ac drive. The nanopatterns are uniaxial nanogrooves with identical and different slope steepness, which induce a pinning potential of the washboard type for Abrikosov vortices. For the fundamental matching field, when the location of vortex rows geometrically matches the nanogrooves, the following effects are observed: The forward transmission coefficient $S_{21}(f)$ of the microstrips can be controllably modulated within a range of about $3$\,dB by the ac current. For the sample with symmetric grooves, depending on the choice of the operation point in the current-voltage curve, the shape and the duty cycle of the output signal can be tuned. For the sample with asymmetric grooves, depending on the ac amplitude, a sine-to-triangular or a sine-to-rectangular pulse shape conversion is observed. The possibility of synthesizing quantized loss levels by a serial connection of the two samples with different nanopatterns is exemplified and can be used for the development of multilevel excess-loss-based fluxonic devices.
\end{abstract}


\maketitle

Superconducting planar transmission lines (SPTLs) with artificial pinning sites offer various possibilities for fundamental research \cite{Plo09tas,Jin10prb,Wor12prb,Sil10inb,Lar15nsr,Luq07prb} and emerging microwave (mw) functionalities \cite{Wal04nat,Hof09nat,Dic09nat,Cla08nat,Son09apl,Bot11apl,Cou13prb,Wor09apl,Dob15apl,Vil03sci}. Examples include circuits for quantum electrodynamics \cite{Wal04nat,Hof09nat} and information processing \cite{Dic09nat}, superconducting qubits \cite{Cla08nat}, resonators \cite{Son09apl,Bot11apl,Cou13prb}, and various \emph{Abrikosov fluxonic devices} \cite{Jin10prb,Sil10inb,Wor09apl,Dob15apl}. The applicabilities of nanopatterned superconductors in passive and active mw devices is determined by the nonlinear vortex dynamics, whereby the crucial role for tailoring the electrical resistance and the mw loss therein is played by the distribution of the pinning sites \cite{Bra95rpp}.

In the last decade, one of the most intensively addressed research lines has been the investigation of the vortex dynamics in the presence of ratchet pinning landscapes \cite{Vil03sci,Plo09tas,Shk14pcm,Shk14ltp,Rei15prb}. These landscapes are characterized by an asymmetry of the pinning potential, that leads to the appearance of a rectified voltage in response to an ac drive. A further tendency in vortex matter research consists in a special interest to studying the vortex dynamics in the regime of GHz frequencies. Namely, from the viewpoint of basic research a few fundamental questions relate to the elucidation of the flux transport mechanisms at microwaves \cite{Wor12prb} and the phenomenon of microwave-stimulated superconductivity \cite{Zol13ltp,Lar15nsr}. From the viewpoint of the development of applications, new functionalities have been reported due to vortex lattice matching effects \cite{Sol14prb,Dob15apl} and the ability to reduce the depinning frequency \cite{Git66prl,Pom08prb,Zai07prb,Jan06prb} by a superimposed dc bias \cite{Dob15apl}. The notion \emph{depinning frequency}, $f_d$, was introduced by Gittleman and Rosenblum in Ref. \cite{Git66prl} who observed that (i) the vortex response is weakly dissipative at low frequencies when the vortices are driven over many pinning sites and it is strongly dissipative at high frequencies when the vortices are each oscillating within one pinning well; (ii) the crossover between these two states can be described on the basis of a mechanistic equation of motion for a single vortex. Accordingly, the depinning frequency has been the key parameter in the theoretical description of the \emph{localization transition} in the vortex dynamics in the radiofrequency and microwave range ever since \cite{Git66prl,Cof91prl,Bra91prl,Zai03prb,Jan06prb,Pom08prb,Shk08prb,Shk11prb,Shk14pcm}. For periodic pinning landscapes it has also been shown by computer simulations \cite{Luq07prb} that for the fundamental matching configuration the vortex-vortex interaction is effectively cancelled so that the dynamics of the entire vortex ensemble can be regarded as that of the \emph{single average vortex} in the \emph{mean pinning potential}.
Recently, it was reported \cite{Sav12prl} that subterahertz transmission of a superconducting metamaterial can be modulated by passing electrical current through it. Quantum metamaterials comprised of networks of superconducting qubits based on Josephson junctions are also used for exploring quantum effects in metaatoms \cite{Ust15met}. Here, we show that in nanopatterned SPTLs the losses due to Abrikosov vortices in the lower GHz range can be modulated and tailored by using an ac transport current, thus allowing SPTLs to enrich the large family of artificial media \cite{Zhe10sci} by the class of \emph{fluxonic metamaterials}. \clearpage

\begin{figure}[b]
    \centering
    \includegraphics[width=0.7\linewidth]{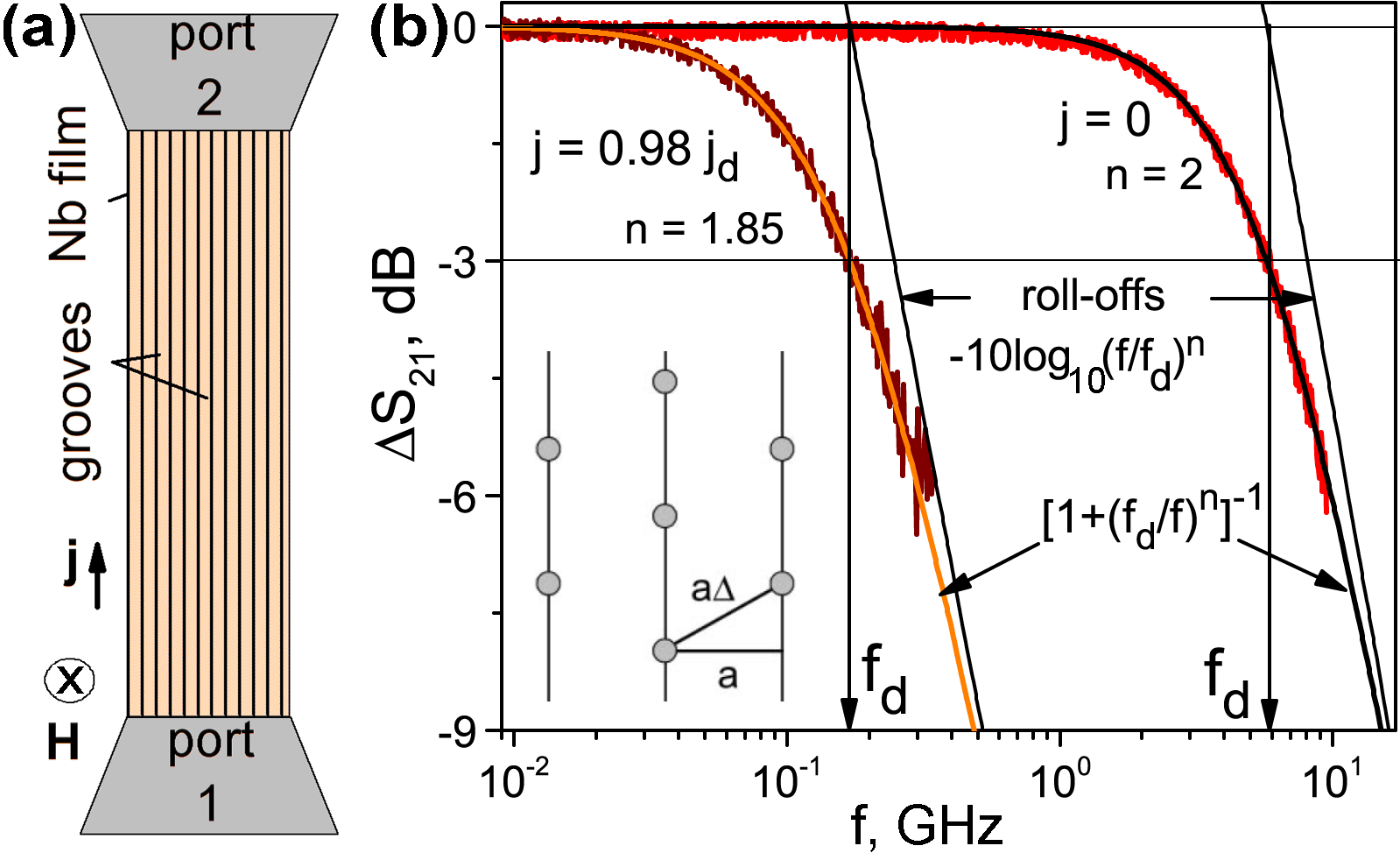}
    \caption{(a)  Experimental geometry. (b) Definition of the depinning frequency $f_d$ by the $-3$\,dB criterion exemplified for a sample with symmetric grooves (sample S). Fits of the frequency characteristic $\Delta S_{21}$ to Eq. (\ref{eDSvF}) are shown by solid lines. The filter roll-offs $-10\log_{10}[(f/f_{d})^n]$ are depicted by the straight lines with the exponents $n$ labeled close to the curves. Inset: The vortex lattice configuration with lattice parameter $a_\bigtriangleup = (2\Phi_0/H\sqrt{3})^{1/2}$ and the matching condition $a_\bigtriangleup = 2a/\sqrt{3}$ in a washboard nanolandscape with period $a$ at the fundamental matching field.}
    \label{SvF}
\end{figure}

\begin{figure}[t!]
    \centering
    \includegraphics[width=0.85\linewidth]{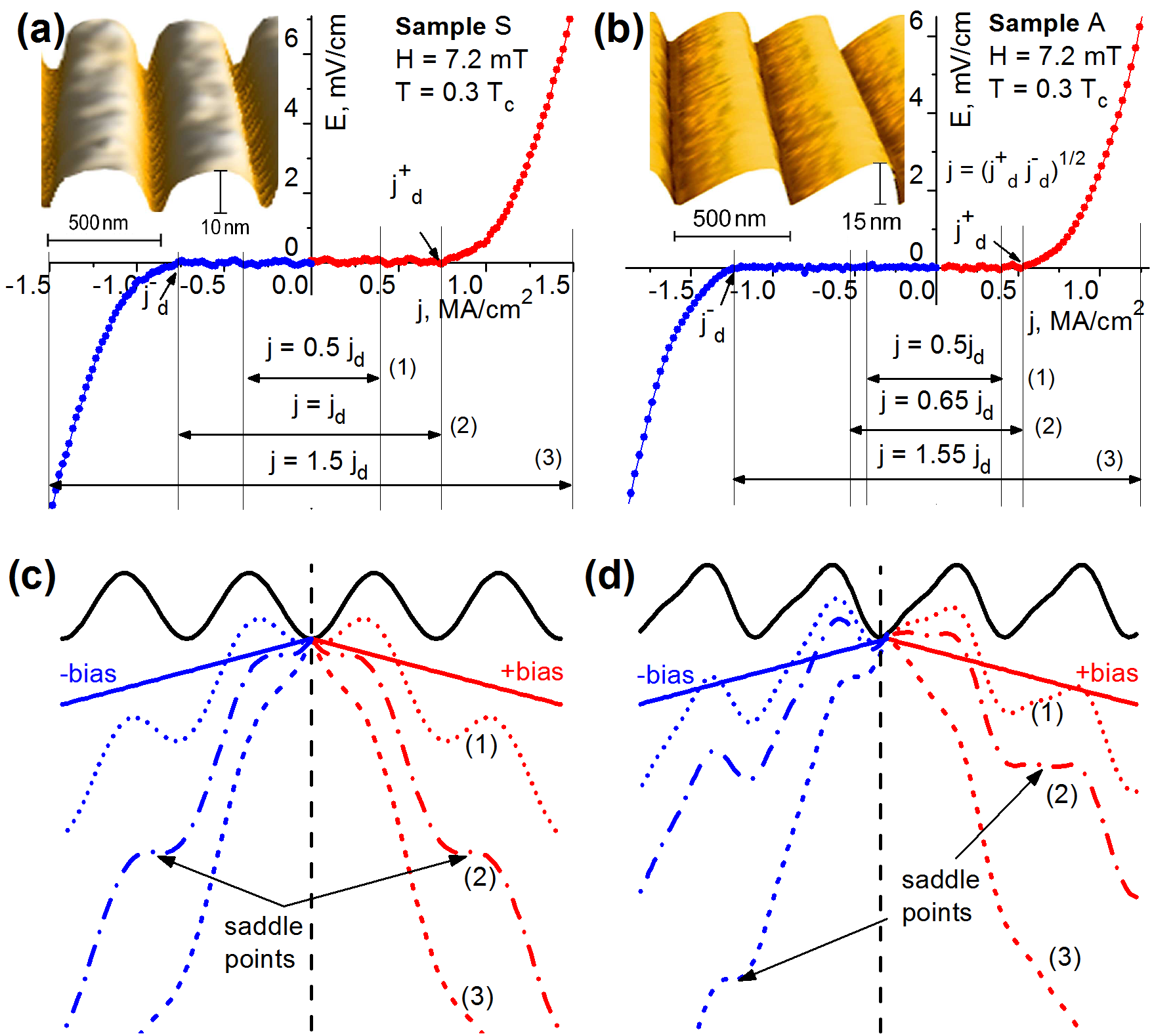}
    \caption{Current-voltage curves for samples S (a) and A (b) at $T = 0.3T_c$ and $H = 7.2$\,mT, respectively. The amplitudes of the quasistatic ac current used for the presentation of the different modulation regimes in Figs. \ref{sampleS} and \ref{sampleA} are shown by the horizontal lines. Tilts of the washboard potentials (c) and (d) sketched for the ac amplitudes (1)---(3) in the CVCs (a) and (b), respectively.}
    \label{CVC}
\end{figure}

Previously, we investigated the frequency dependence of the mw power absorbed by vortices in Nb microstrips with symmetric and asymmetric pinning nanostructures of the washboard type in small perpendicular magnetic fields \cite{Dob15apl}, see Fig. \ref{SvF}(a) for geometry. The absorbed mw power was measured in terms of the absolute value of relative change of the forward transmission coefficient $\Delta S_{21}(f) \equiv S_{21} - S_{21\mathrm{ref}}$, where $S_{21\mathrm{ref}}$ stands for the reference mw loss in the transmission line (all cables, connectors etc.) and, hence, $\Delta S_{21}$ provides a measure for the mw loss due to vortex motion in the sample under study. We found \cite{Dob15apl} that for the fundamental vortex lattice matching configuration  with the pinning nanolandscape (see inset to Fig. \ref{SvF}(b), $\Delta S_{21}(f)$ has a smooth step-like functional shape, as shown in Fig. \ref{SvF}(b). One of the key observations of that work was that the depinning frequency defined at the $-3$\,dB excess loss level is shifted towards low frequencies upon increasing the dc bias value. The frequency dependence of the mw loss in both microstrips was fitted well to the phenomenological expression
\begin{equation}
    \label{eDSvF}
    \begin{array}{lll}
    \Delta S_{21}(f) = 1/[1 + (f_{d}/f)^n],\qquad\qquad |j|< |j_d|,\\[3mm]
    \Delta S_{21}(f) \approx -6.4\,\mathrm{dB},\qquad\qquad\qquad\quad |j|> |j_d|,
    \end{array}
\end{equation}
with the exponent $n=2$ for dc biases $j$ not very close to the depinning current $j_d$ and $n\approx 1.9$ for $j\rightarrow j_d$. Here, the depinning current is determined from the current-voltage curves (CVC) by the $0.1$\,$\mu$V/cm electric field strength criterion. In the mechanistic scenario of  vortex motion in a tilted potential \cite{Shk08prb,Shk11prb,Shk14pcm} this physically corresponds to the disappearance of the pinning potential barrier, refer to Fig. \ref{CVC}(c) and (d). Accordingly, in Eq. (\ref{eDSvF}) the excess loss level $-6.4$\,dB corresponds to the nearly linear flux-flow regime in the CVC. Equation (\ref{eDSvF}) corresponds to a first-order filter \cite{Poz11boo} roll-off $-10\log_{10}[(f/f_d)^2]$ related to the frequency characteristics of the microstrip as shown in Fig. \ref{SvF}. A crucial observation \cite{Dob15apl} for the subsequent presentation is that the depinning frequency of the microstrip with an asymmetric pinning landscape depends not only on the dc bias value but also on its polarity.

Here, we superimpose a low-frequency ($3$\,Hz) ac current on the mw stimulus and provide evidence that the ac drive can be straightforwardly used for the modulation and synthesis of the mw excess loss levels by vortices. The samples are the same as in Ref. \cite{Dob15apl}. These are two $150\times500\,\mu$m$^2$ microstrips made from epitaxial (110) Nb films sputtered onto heated a-cut sapphire substrates \cite{Dob12tsf}. The nanopatterns are uniaxial grooves with a period $a=500$\,nm [see insets to Fig. \ref{CVC}(a) and (b)] fabricated by focused ion beam (FIB) milling \cite{Dob12njp}. The grooves are parallel to the direction of the transport current density $\mathbf{j}$ and, hence, in a perpendicular magnetic field $\mathbf{H}$ the dc dissipation and the mw loss in the microstrip are related to the vortex dynamics across the grooves. One microstrip was patterned with grooves having a symmetric cross-section (sample S) and the other with grooves having an asymmetric one (sample A). Samples S and A are 40\,nm and 70\,nm thick and have a critical temperature $T_c$ of $8.66$\,K and $8.94$\,K, respectively. The upper critical field for both samples at zero temperature $H_{c2}(0)$ is about $1$\,T as deduced from fitting the dependence $H_{c2}(T)$ to the phenomenological law $H_{c2}(T) = H_{c2}(0) [1-(T/T_c)^2]$.

Combined broadband mw and dc electrical resistance measurements were done in a $^4$He cryostat with a custom-made sample probe \cite{Dob15mst} at the temperature $T = 2.65$\,K $\approx 0.3T_c$ for both samples. The mw signal was fed to the sample through coaxial cables from an Agilent E5071C vector network analyzer (VNA, $300$\,KHz -- $14$\,GHz). The quasistatic ac current was superimposed and uncoupled by using two bias-tees mounted at the VNA ports. For all frequencies, the mw excitation power at the sample is $P = - 20$\,dBm ($10\,\mu$W) as kept by the VNA in accordance with the pre-saved calibration data. In the absence of ac current, at $H=7.2$\,mT and $T = 0.3T_c$ microstrips S and A are characterized by the depinning frequencies  $f_d = 5.72$\,GHz and $f_d = 3.02$\,GHz, respectively \cite{FdepAPL}. The depinning frequencies have been chosen as  \emph{carrier frequencies} for the mw signal for which the excess losses modulated by the ac current are discussed next.\clearpage

The CVCs of the microstrips measured for \emph{dc currents} of both polarities are shown in Fig. \ref{CVC}(a) and (b). While the CVC for sample S has equal depinning currents in both branches, one clearly sees the difference in the depinning currents for the positive $j^+_d$ and the negative $j^-_d$ polarity for sample A. Here and in all the following figures the positive ac halfwaves and the related responses are plotted in \emph{red}, while the negative ac halfwave and the related responses are shown in \emph{blue}. The difference in $j_d$ under current direction reversal is the fingerprint of a vortex ratchet system \cite{Plo09tas,Sil10inb,Shk14pcm}. Each of panels (c) and (d) in Fig. \ref{CVC} reports \emph{different} snapshots of the potential tilts for \emph{positive and negative} dc biases, respectively. For a dc current of one and the same polarity the respective modification of a symmetric potential was reported in \cite{Shk11prb}.

In the next experiment with quasistatic ac current $j \equiv j(t) = j\sin\omega t$ with $\omega = 2\pi/t_f$ we consider three different regimes depending on the relation of the ac amplitude to the depinning currents for each sample, as denoted by the horizontal lines in the bottom parts of Fig. \ref{CVC}(a) and (b). Namely, these regimes correspond to the case of subcritical $j/j_d = 0.5$, critical  $j/j_d = 1$ and overcritical $j/j_d = 1.5$ current for sample S, see regimes (1), (2) and (3) in Fig. \ref{CVC}(a) and the respective tilts of the pinning potential in Fig. \ref{CVC}(c). To present only the qualitatively different and most interesting regimes for sample A, in Fig. \ref{CVC}(b) and (d) we consider the case $j/j_d = 0.5$ as representative for subdepinning currents for both CVC branches, $j/j_d = 0.65$ corresponding to the weak-slope depinning current in the positive CVC branch, and the case $j/j_d = 1.55$ corresponding to the strong-slope depinning current in the negative CVC branch, curves (1)---(3) in Fig. \ref{CVC}(b) and (d), respectively. We also underline that the \emph{ratchet window} of the system ensue for the current range $|j^+_d |< |j| < |j^-_d|$ \cite{Shk14pcm,Shk14ltp}. For both samples, the depinning current density is defined as $j_d = (j_d^+ j^-_d)^{1/2}$, where $j^+_d = 0.52$\,MA/cm$^2$ and $j^-_d = 1.25$\,MA/cm$^2$ for sample A, while $j^+_d = j^-_d = 0.75$\,MA/cm$^2$ for sample S.

\begin{figure}[t!]
    \centering
    \includegraphics[width=0.9\linewidth]{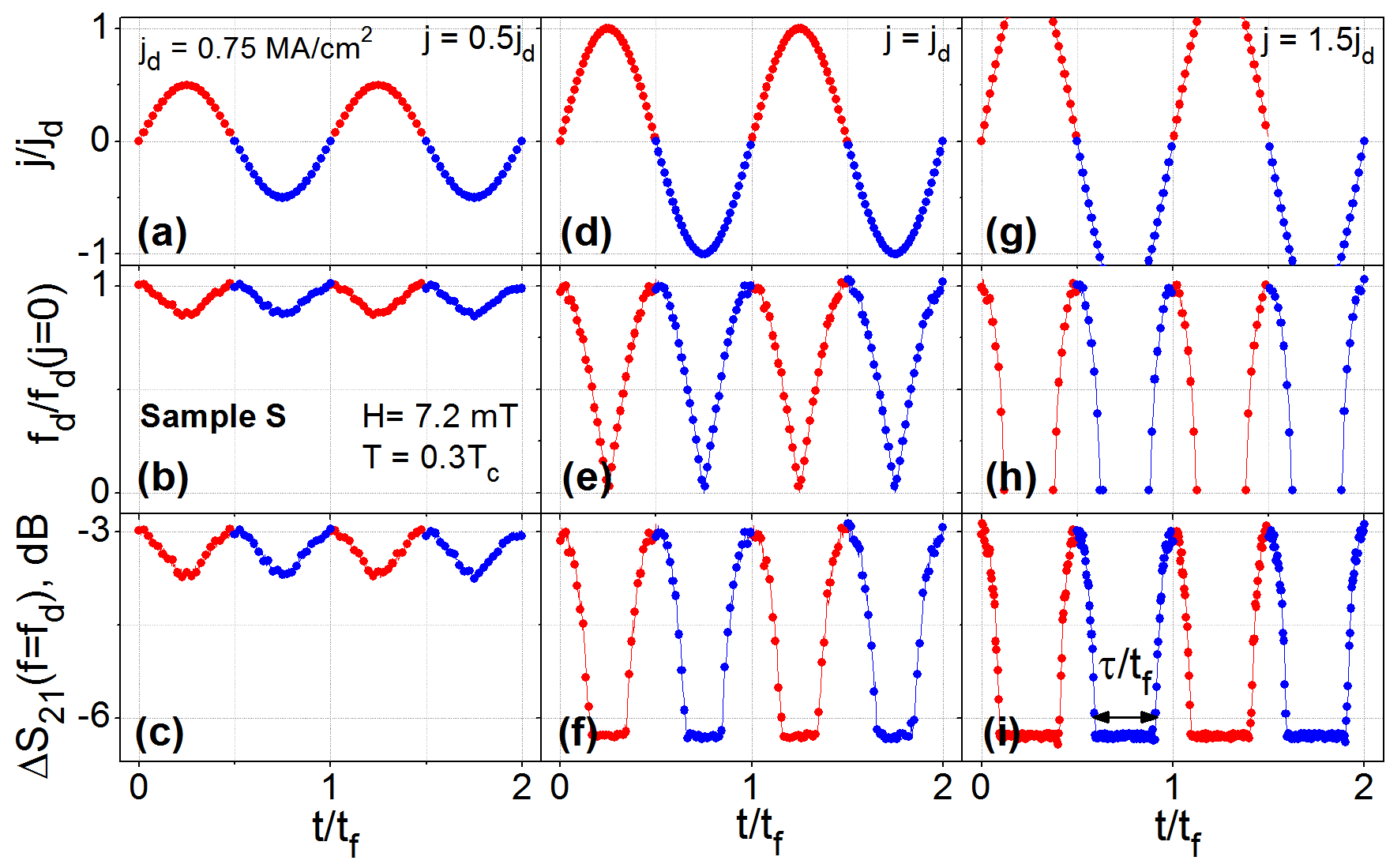}
     \caption{Microwave loss modulation in microstrip S at $H = 7.2$\,mT, $T = 0.3T_c$, and the excitation power $P = - 20$\,dBm. The adiabatic ac current ($3$\,Hz) in the subcritical (a), close-to-critical (d) and overcritical (g) regime leads to the reduction of the depinning frequency [panels (b), (e), and (h), respectively] and the appearance of an excess loss at $5.72$\,GHz due to the vortices $\Delta S_{21}(f=f_d(j=0))$ [panels (c), (f), and (i), respectively]. The rectangular ``floors'' in panels (f) and (i) correspond to the maximal loss value in the flux-flow regime.}
    \label{sampleS}
\end{figure}

\begin{figure}[t!]
    \centering
    \includegraphics[width=0.9\linewidth]{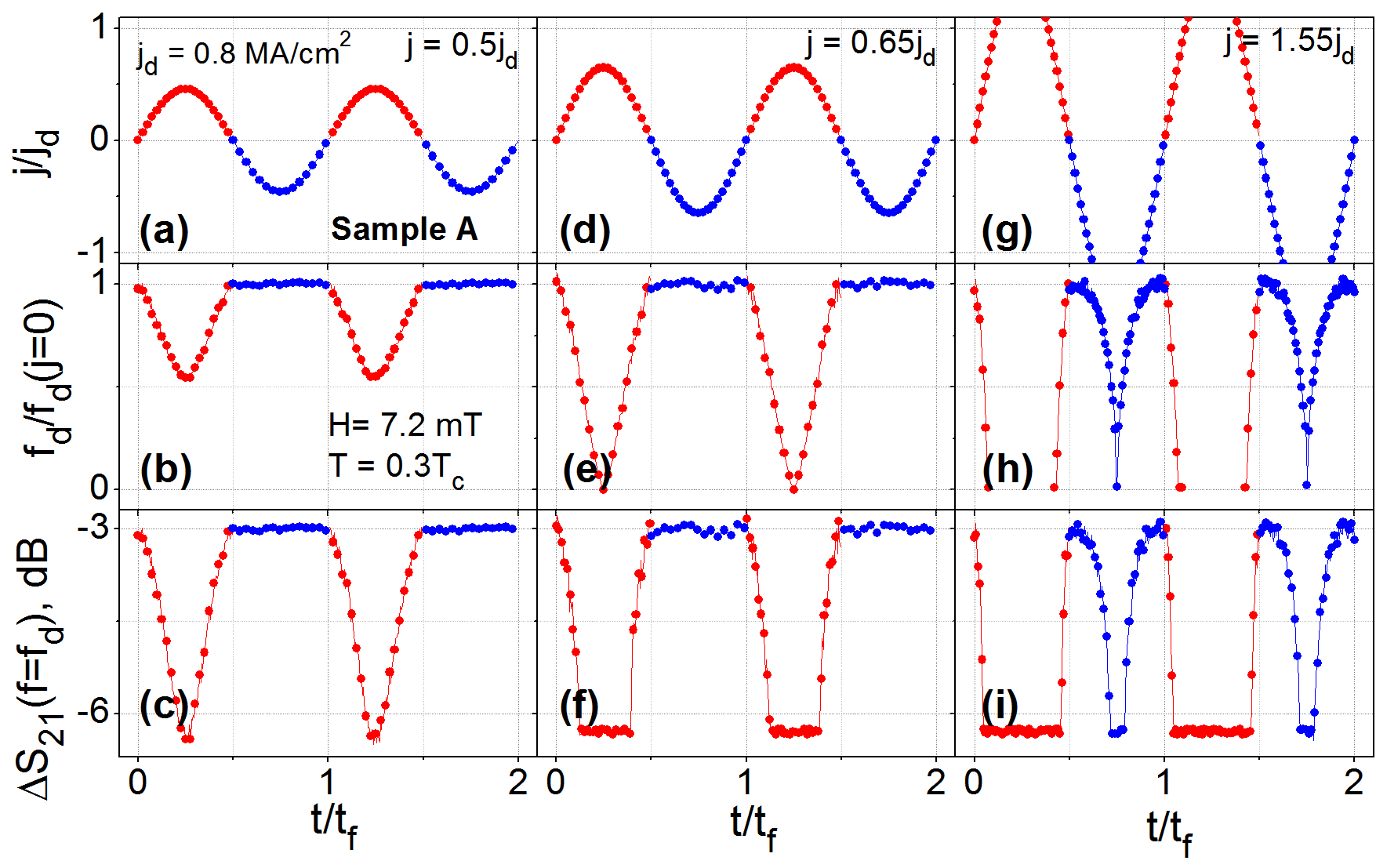}
     \caption{Microwave loss modulation in microstrip A at $H = 7.2$\,mT, $T = 0.3T_c$, and the excitation power $P = - 20$\,dBm. The adiabatic ac current ($3$\,Hz) in the subcritical (a), close-to-critical (d) and overcritical (g) regime leads to the reduction of the depinning frequency [panels (b), (e), and (h), respectively] and the appearance of the excess loss at $3.02$\,GHz due to the vortices $\Delta S_{21}(f=f_d(j=0))$ [panels (c), (f), and (i), respectively].}
    \label{sampleA}
\end{figure}

The sine waves of the input quasistatic currents are shown in the upper row of the panels in Figs. \ref{sampleS} and \ref{sampleA}. The reduction of the depinning frequency upon ramping the current value is shown in the middle panels of the respective figures. The panels in the bottom row in Figs. \ref{sampleS} and \ref{sampleA} present the time dependences of the excess loss $\Delta S_{21}(t, f=f_d)$. In particular, in Fig. \ref{sampleS}(b) one sees that in the regime of subcritical ac amplitudes, the modulation of the depinning frequency with respect to its value in the absence of the quasistatic current is rather small, which results in a smooth modulation of $\Delta S_{21}(t)$ within about $1$\,dB. The modulation of $f_d$ attains its maximal depth with a module-of-sine shape for the critical-amplitude regime in Fig. \ref{sampleS}(e). Accordingly, $\Delta S_{21}$ exhibits  ``floors'' at $-6.4$\,dB corresponding to the maximum excess loss level due to the vortex motion in the flux-flow regime. We recall that the measurements are taken at the mw frequency $f=f_d$, that is in the most sensitive point of $\Delta S_{21}(f)$. For this reason even though the depinning current is not reached a transition to the strongly dissipative state in the vortex response takes place. With further increase of the ac amplitude in Fig. \ref{sampleS} (g), the curves $f_d(t)$ in Fig. \ref{sampleS} (h) are characterized by gaps owing to that the depinning frequency loses its physical meaning for the flux-flow regime of vortex motion. The respective  $\Delta S_{21}(t)$ traces demonstrate a long duty cycle for the excessively lossy state. It should be noted that there is no difference in the modulated signal for the positive and the negative halfwaves for sample S. By the appropriate choice of the operating point in the CVC, the duty cycle, defined as $\tau/t_f$ at the $-6$\,dB level, can be tuned from about $5\%$ to $95\%$.\clearpage

The modulation patterns for sample A are substantially different. Namely, in regime (1) in Fig. \ref{CVC}(b) the depinning frequency is modulated during the \emph{positive ac halfwave only}, whereas it remains constant during the negative ac halfperiod, see Fig. \ref{sampleA}(b). In consequence of this, the excess loss $\Delta S_{21}(t)$ is observed during the positive ac halfwaves and is absent during the negative one. Interestingly, the time dependence of $\Delta S_{21}(t)$, where it is non-constant, has a nearly-triangular, saw-tooth form, i.\,e., a \emph{sine-to-triangular pulse form conversion} takes place. We attribute this effect to the particular form of the pinning potential induced by the asymmetric landscape in sample A. When the weak-slope depinning current is reached in regime (2) in Fig. \ref{CVC}(b), the depinning frequency is suppressed to its minimal value during the positive ac halfwave and remains constant during the negative ac halfwave, see Fig. \ref{sampleA}(e). It is worth noting that the shape of the $\Delta S_{21}(t)$ curve in this case is nearly rectangular, that is a \emph{sine-to-rectangular pulse form conversion} takes place. It is this regime which represents the \emph{microwave loss analogue} of the ratchet effect that is well known for the dc voltage \cite{Plo09tas}. Finally, in the regime of strong-slope critical drive in Fig. \ref{sampleA}(i), the excess loss $\Delta S_{21}(t)$ combines the features of the two previous regimes for the positive and negative ac halfwaves. Hence, microstrip A demonstrates a signature of a \emph{waveform synthesizer} which will be examined in a control experiment described in the last but one paragraph.

Given the adiabatic current $j = j \cos 2\pi t/t_f$ and Eq. (\ref{eDSvF}) for the relation between the mw loss and the depinning frequency, the time dependences $\Delta S_{21}(t, f =f_d)$ allow one to approximate the reduction of the depinning frequency for both CVC branches in Fig. \ref{CVC}(a) for sample S by the following expression:
\begin{equation}
    \label{eDFvJS}
    f_d/f_d(j =0) = [1 - (j/j_d)^2]^{1/2},\qquad |j|< |j_d|.\\
\end{equation}
At the same time, for the gentle-slope direction of the asymmetric potential of sample A [positive CVC branch in Fig. \ref{CVC}(b)], probed by the positive halfwave of the ac current, the fit reads
\begin{equation}
    \label{eDFvJAsteep}
     f_d/f_d(j =0) = [1 - (j/j_d)^{3/2}]^{2/3},\qquad 0 < j< j^+_d,
\end{equation}
while for its steep-slope direction [negative CVC branch in Fig. \ref{CVC}(b)], probed by the positive halfwave of the ac current, the dependence reads
\begin{equation}
    \label{eDFvJAsteep}
     f_d/f_d(j =0) = [1 - (j/j_d)^4]^{1/4},\qquad -j^-_d  < -j< 0.
\end{equation}
The dependences by Eqs. (\ref{eDFvJS})--(\ref{eDFvJAsteep}) in conjunction with Eq. (\ref{eDSvF}) represent the transfer functions of the microstrips, thus allowing one to \emph{analytically describe and experimentally tailor the modulation of the microwave losses} therein.

\begin{figure}[b!]
    \centering
    \includegraphics[width=0.55\linewidth]{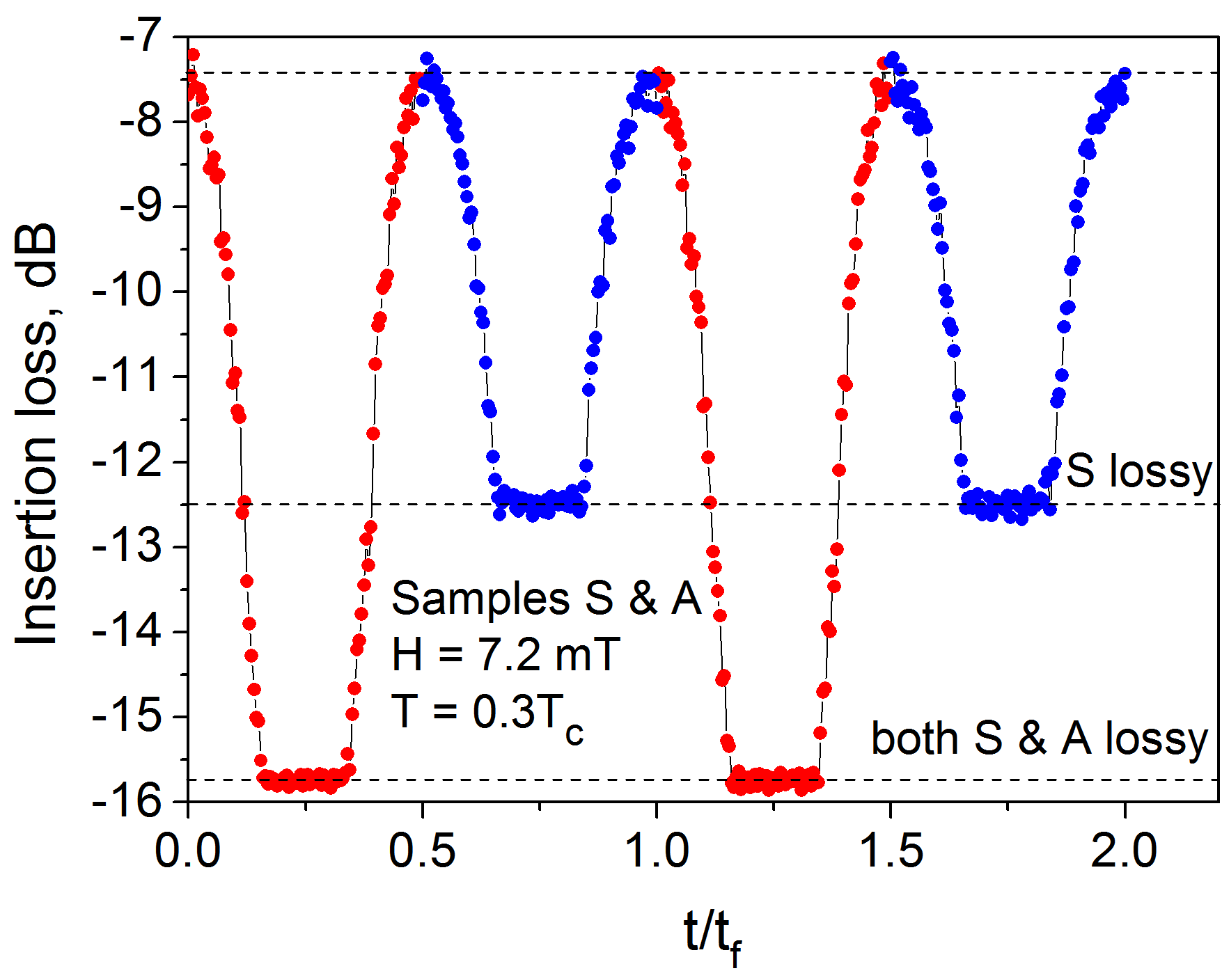}
     \caption{Example of synthesizing (a) three different insertion loss levels by a serial connection of samples S and A for an ac current $I = 50$\,mA and a frequency $3.02$\,GHz.}
    \label{IL}
\end{figure}

Finally, to further elaborate on the excess loss level synthesis, we consider the mw transmission through samples S and A connected in series. For definiteness, we apply the ac current $I$ with an amplitude of $50$\,mA that, given the thicknesses of the samples, corresponds to the current density amplitude $0.48$\,MA/cm$^2$ for sample A and $0.83$\,MA/cm$^2$ for sample S. We note that $j=0.48$\,MA/cm$^2 \simeq 0.52$\,MA/cm$^2=j_d^+$ for sample A and $j=0.83$\,MA/cm$^2 \simeq 0.75$\,MA/cm$^2=j_d$ for sample S. Accordingly, the cumulative insertion loss in Fig. \ref{IL} is characterized by three different insertion loss levels, whereby the intermediate level is achieved in consequence of the combination of the lossy state of microstrip S and the low-lossy state of microstrip A. This can be understood as a result of summation of the individual excess loss levels in samples S and A with the help of panels (f) in Figs. \ref{sampleS} and \ref{sampleA}, respectively, given the larger difference (about $5$\,dB) in $\Delta S_{21}$ for sample S at $3.02$\,GHz  in Fig. \ref{SvF} with respect to the maximum excess loss level $-6.4$\,dB. Therefore, this approach appears feasible also for a larger number of differently-patterned microstrip sections composing a \emph{fluxonic metamaterial}.

In summary, we have introduced afluxonic metamaterial on the basis of superconducting Nb microstrips with symmetric and asymmetric nanogrooves and investigated the time dependences of the excess losses $\Delta S_{21}(t)$ due to Abrikosov vortices in the presence of a quasistatic ac current for the fundamental geometrical matching of the vortex lattice with the periodic pinning nanolandscape ($H = 7.2$\,mT). The ac-driven modulation of $\Delta S_{21}(t)$ accompanied by a \emph{sine-to-triangular and a sine-to-rectangular pulse shape conversion} has been reported and the possibility of \emph{synthesizing pre-defined quantized loss levels} for serial connection of samples S and A has been exemplified. These findings are relevant for the development of more sophisticated, multilevel excess-loss-based fluxonic devices. The reported effects become entangled and eventually vanish upon detuning the field value away from 7.2\,mT. The analysis of the microwave loss in the work was done for two carrier frequencies corresponding to the vortex depinning frequencies for samples S and A amounting to $5.72$\,GHz and $3.02$\,GHz, respectively, as defined by the $-3$\,dB excess loss criterion. If, however, a deeper modulation of the microwave losses is desired, the carrier frequency should be halved so that the ultimate modulation depth of about $6$\,dB can be achieved.

OVD thanks Roland Sachser for automating the data acquisition and helping with the nanopatterning. This work was financially supported by the German Research Foundation (DFG) through grant DO 1511/2-4 and conducted within the framework of the NanoSC-COST Action MP1201 of the European Cooperation in Science and Technology.


\begin{thebibliography}{39}
\bibitem [{Fde()}]{FdepAPL}%
  \BibitemOpen
  \href@noop {} {}\bibinfo {note} {The difference in $f_d$ is attributed to the
  different film thicknesses $d$, whereby $f_d(d)$ has been found to obey the
  phenomenological law $f_d\propto1/d$ \cite{Dob15apl}. With increasing
  temperature and magnetic field the depinning frequency is reduced and can be
  approximated by the expressions $f_d(T) = f_d(0)[1-(T/T_c)^4]$
  \cite{Zai03prb} and $f_d(H) = f_d(0)[1-(H/H_{c2})^2]$ \cite{Jan06prb}, where
  $f_d(0,0)$ is the depinning frequency at $T=0$ and $H=0$.}\BibitemShut
  {Stop}%
\end{thebibliography}
\end{document}